\documentclass[aps,prl,twocolumn,preprintnumbers,showpacs,superscriptaddress,nofootinbib,floatfix,10pt]{revtex4-1}

\pdfoutput=1

\usepackage{textcomp}
\usepackage[english]{babel}
\usepackage{amsmath,amssymb,amsbsy,booktabs}
\usepackage{bm}
\usepackage{color}
\usepackage{array}
\usepackage{amstext}
\usepackage{graphicx}
\usepackage{amsfonts}
\usepackage{bm}
\usepackage{dcolumn}
\usepackage{rotating}
\usepackage{epstopdf}
\usepackage{esint}
\usepackage{url}
\usepackage[colorlinks=true,
linkcolor=blue,
filecolor=blue,
anchorcolor=blue,
urlcolor=blue,
citecolor=blue
]{hyperref}

\newcommand{\TT}[1]{{\bf {#1}}}

\makeatletter
\newcommand{\thickhline}{%
	\noalign {\ifnum 0=`}\fi \hrule height 1pt
	\futurelet \reserved@a \@xhline
}
\makeatother
\allowdisplaybreaks

\begin{document}

	\title {\bf Freeze-in self-interacting dark matter in warped extra dimension}
	\author{Shao-Ping Li}
	\email{ShowpingLee@mails.ccnu.edu.cn}
	\affiliation{Institute of Particle Physics and Key Laboratory of Quark and Lepton Physics~(MOE),\\
		Central China Normal University, Wuhan, Hubei 430079, China}
	
	\author{Xin Zhang}
	\email{xinzhang@hubu.edu.cn}
	\affiliation{Faculty of Physics and Electronic Science, Hubei University, Wuhan, Hubei 430062, China}

\begin{abstract}
A classically scale-invariant scalar singlet can be  a MeV-scale dark matter, with a feeble Higgs portal coupling at $\mathcal{O}(10^{-10})$. Besides, an $\mathcal{O}(0.1)$   self-interaction coupling   could further serve to alleviate the small-scale problems in the Universe. We show  that, such a  dark matter candidate can naturally arise in the warped extra dimension,  with the   huge span of parameter space   predicted well within $\mathcal{O}(1)$   fundamental parameters.
	
\end{abstract}


\maketitle

\TT{Introduction:} Warped extra-dimension (WED) theory  in the Randall-Sundrum (RS) model has shown   a glorious insight to solve the hierarchy problem~\cite{Randall:1999ee}.  The original RS solution to the large span between the Planck and electroweak scales set  the Standard Model (SM) Higgs field to be  brane-localized  on  a boundary point of the compactified extra dimension (the so-called IR or TeV brane),  in which all the SM particles also populate. Later  it was found that,  if the fermions  propagate in the fifth dimension, the feebleness and hierarchies of Yukawa couplings can also be addressed in a \textit{natural} way~\cite{Grossman:1999ra,Huber:2000ie,Gherghetta:2000qt}. WED subsequently  becomes a common stilt to explain hierarchies of couplings and scales existed in various phenomenological models.

The successful freeze-in production of  dark matter (DM) with a feeble portal coupling   has been studied intensively  in recent years (see, e.g., refs.~\cite{Hall:2009bx,Yaguna:2011qn,Blennow:2013jba,Heikinheimo:2016yds,Heeba:2018wtf}). Depending on the specification of DM mass, an $\mathcal{O}(10^{-12}) -\mathcal{O}(10^{-9})$ portal coupling  is generically required to fit the DM relic abundance. Among the  possible freeze-in DM models, a real scalar  singlet $S$ is arguably the simplest candidate, which    interacts with  the SM    via the quartic Higgs portal $H^\dagger H S^2$.
While such a framework   is quite simple,
the question about the origin of  feeble portal coupling remains to be solved. Thus far, this   has stirred up an increased attention on sourcing the tiny portal~\cite{Elahi:2014fsa,Kim:2017mtc,Kim:2018xsp,Borah:2018gjk,Goudelis:2018xqi}.

In the real scalar singlet models, it was found earlier in ref.~\cite{McDonald:2001vt} that, if the DM mass stems entirely from the Higgs vacuum, i.e., the DM has a classical scale-invariance, its mass and relic density would share a common origin~\cite{Kang:2015aqa}.  In addition,  if  the DM self-interaction coupling resides at $\mathcal{O}(0.1)$, an  order similar to the SM Higgs  quartic coupling,  the self-interacting DM could further alleviate the  small-scale discrepancies between the  collisionless cold DM simulations and the structure observations in the Universe~\cite{McDonald:2001vt} (see  ref.~\cite{Spergel:1999mh} for the original proposal, as well as   the recent review~\cite{Tulin:2017ara} and references therein). This simple DM model, however,  remains a huge coupling span ranging from $\mathcal{O}(0.1)$ to $\mathcal{O}(10^{-10})$, and it seems that the explanation of DM relic density and small-scale problems relies on an unnaturally numerical coincidence.

 We will show in this paper that, the classically scale-invariant DM   can arise  in the WED as the zero-mode Kaluza-Klein (KK)  scalar.  The DM production can be approximately ascribed to a two-step freeze-in mechanism. The zero-mode profile induced from the classical  scale-invariance plays the key role to    generate  the  numerical coincidence mentioned above,  from which the  huge span of parameter space  is   unified  via natural $\mathcal{O}(1)$ couplings in the WED.  In the following, we begin by embedding a scale-invariant scalar singlet in the WED with a five-dimensional (5D)  self-interaction and a Higgs portal localized on the IR brane,   then we apply the simple freeze-in mechanism with a two-step production before and after the electroweak symmetry breaking (EWSB) under some proper approximations, and calculate the DM relic density as well as the ratio of cross section over mass before we finally conclude.

\TT{Scale-invariant singlet in the WED:} The starting point to consider  the RS model~\cite{Randall:1999ee} is featured by the 5D metric
\begin{align}
	ds^2=e^{-2\sigma(\phi)} \eta_{\mu\nu} dx^\mu dx^\nu-r_c^2 d\phi^2,
\end{align}
where the fifth dimension  is compactified on an $S^1/Z_2$ orbifold, with the compactification radius $r_c$ and  angular coordinate $\phi\in [-\pi, \pi]$. The $Z_2$ symmetry on the orbifold dictates  a $\phi$-periodicity where $(x,\phi)$ is identified with $(x,-\phi)$. $e^{-2\sigma(\phi)}$ is the warped factor,  with $\sigma(\phi)\equiv k r_c \vert \phi \vert$.
The induced  4D Planck mass is related to the 5D fundamental scale $M$ via $M_{Pl}^2=M^3(1-e^{-2 k r_c \pi})/k$, with    the curvature $k$ of the fifth dimension at $k\simeq M\simeq 10^{19}$~GeV.

If the SM Higgs sector  is localized on the $\phi=\pi$ brane, i.e., the so-called  IR (TeV) brane, from which the action is constructed  as
\begin{align}
	S_{H}=\int d^4 x \sqrt{\tilde{g}} \left(\tilde{g}^{\mu\nu} D_\mu H^\dagger D_\nu H-\lambda (\vert H\vert^2-v_0^2)^2\right),
\end{align}
where  $\tilde{g}_{\mu\nu}=e^{-2k r_c \pi} g_{\mu\nu}$, with the flat limit of $g_{\mu\nu}$ given by the  Minkowskian metric signature $\eta_{\mu\nu}=(1,-1,-1,-1)$.
The canonical normalization   is realized by  Higgs wavefunction rescaling, $H\to e^{k r_c \pi} H$, which makes the  EWSB scale $v_0$  shifted to the physical one via $v=e^{-k r_c \pi}v_0$.
It can then be found that, with $k r_c\simeq 12$,  the electroweak vacuum $v\simeq 246$~GeV is  induced from the non-hierarchical scale $v_0\simeq M$ in the fundamental 5D regime, and hence the  hierarchy problem  can  be explained.

In the 5D perspective,    a natural suppression of Higgs portal coupling implies a bulk scalar DM    propagating  in the fifth dimension with an exponential-like profile.   Let us
consider the DM scale   solely generated by the electroweak vacuum~\cite{McDonald:2001vt}. In this case,  the 4D scalar potential would be of the form
\begin{align}\label{4Dpotential}
	V(S,H)=-\mu_{H}^2 \vert H\vert^2+\frac{\lambda_H}{4} \vert H\vert^4+\lambda_P S^2 \vert H\vert^2 +\frac{\lambda_S}{4} S^4,
\end{align}
where the EWSB occurs  via the Higgs mechanism.
To  induce the above potential in the 5D framework,  we propose a scalar action with a brane-localized Higgs portal: $	S_{scalar}=S_{\Phi}+S_{H}$,  in which
\begin{align}\label{saction}
	S_{\Phi}=\int d^4 x \; d\phi \;e^{-4\sigma}r_c&\Big ( \frac{1}{2}G^{AB}\partial_A \Phi \partial_B \Phi -\frac{1}{2}M^2_{\Phi}\Phi^2
	\nonumber \\
	&+\frac{\lambda_S^{(5)}}{4}\Phi^4
	+\lambda_{P}^{(5)} \Phi^2 \vert H\vert^2\delta_\pi\Big ),
\end{align}
with  $G^{\mu\nu}=e^{2\sigma}g^{\mu\nu}, G^{\phi \phi}=-1/r_c^2$, $\delta_\pi\equiv \delta(\phi-\pi)$, and a 5D bare mass term~\cite{Gherghetta:2000qt}:  $M_\Phi^2\equiv a k^2+b\sigma^{\prime\prime}(\phi)/r_c$,
where $a,b\simeq \mathcal{O}(1)$ are dimensionless parameters, and
\begin{align}
\sigma^{\prime\prime}(\phi)=2k (\delta(\phi)-\delta(\phi-\pi)).
\end{align}
Applying the KK decomposition
\begin{align}		
	\Phi(x,\phi)=\sum_n S_n(x) \frac{f_n(\phi)}{\sqrt{r_c}},
\end{align}
we can see that, the 5D  portal coupling $\lambda_{P}^{(5)}\simeq \mathcal{O}(1)$ is dimensionless while $\lambda_{S}^{(5)}$ has dimension $-1$, which, without loss of generality,  can be parametrized as $\lambda_{S}^{(5)}\equiv \xi/k$, with $\xi\simeq \mathcal{O}(1)$.

The equation of motion (EoM) for the free scalar profile  has been obtained earlier in ref.~\cite{Goldberger:1999wh}, which reads
\begin{align}\label{sEoM}
	\frac{1}{r_c^2}\partial_\phi\left(e^{-4\sigma(\phi)}\partial_\phi f_n \right)-M_\Phi^2 e^{-4\sigma(\phi)}f_n=-m_n^2e^{-2\sigma(\phi)}f_n.
\end{align}
 For the zero mode with a classical scale invariance, we have $m_0=0$.
Solving eq.~\eqref{sEoM} with the following  boundary conditions~\cite{Gherghetta:2000qt}
\begin{align}
 f_n^\prime(\phi)-b \sigma^\prime f_n(\phi)\Big|_{\phi=0,\pi}=0,
\end{align}
 we obtain two possible solutions for the zero mode
\begin{align}
	b=\pm\sqrt{4+a}+2.
\end{align}
Given   the orthonormal condition
\begin{align}\label{snorm}
	\int^\pi_{-\pi}d\phi e^{-2\sigma(\phi) }f_n(\phi)f_m(\phi)=\delta_{mn},
\end{align}
it is easy to check that, only the portal coupling  obtained in the $b=-\sqrt{4+a}+2$ direction can have an exponential-like scaling, corresponding to a normalized zero-mode profile
\begin{align}
	f_0(\phi)= \sqrt{\frac{kr_c (\sqrt{a+4}-1)}{1-e^{-2k r_c\pi (\sqrt{a+4}-1)}}}e^{- k r_c(\sqrt{a+4}-2)\vert \phi\vert}.
\end{align}
The induced 4D portal coupling  $\lambda_{P,0}$ and quartic coupling $\lambda_{S,0}$  of the zero mode are given respectively by
\begin{align}
	\lambda_{P,0}&=\lambda_{P}^{(5)} e^{-2\sigma(\pi)}f_0^2(\pi)
		\nonumber \\[0.1cm]
	&\simeq \lambda_P^{(5)}k r_c(\sqrt{a+4}-1)e^{-2kr_c\pi (\sqrt{a+4}-1)},\label{lambdaP}
 \\[0.2cm]
	\lambda_{S,0}&=\frac{\xi}{k r_c} \int^\pi_{-\pi} d\phi e^{-4\sigma(\phi)} f_0^4(\phi),
	\nonumber \\[0.1cm]
	&\simeq \frac{1}{2}\xi(\sqrt{a+4}-1) \coth[kr_c\pi(\sqrt{a+4}-1)].\label{lambdaS}
	\end{align}
It can readily be   seen that, $\lambda_{P,0}$ is exponentially suppressed provided that $ a\geqslant -3$. Besides, the hyperbolic cotangent  indicates that the quartic coupling $\lambda_{S,0}$ would  be a stable function of parameter $a$ even though it controls the exponential scaling of portal coupling $\lambda_{P,0}$.

 Some remarks are made here about the EoM in eq.~\eqref{sEoM}.  The identification of 4D KK scalar mass $m_n$ can be obtained from the 5D EoM, which, after including the self-interaction in the bulk, gives
	\begin{align}
		\frac{1}{\sqrt{G}}\partial_A(\sqrt{G}G^{AB}\partial_B \Phi)+M_\Phi^2\Phi-\lambda_S^{(5)}\Phi^3=0,
	\end{align}
or in terms of the KK decomposition, reads
	\begin{align}\label{5DEoM}
	0&=\sum_n e^{2\sigma} f_n g^{\mu\nu} \partial_\mu \partial_\nu S_n
	\nonumber \\
	&-\sum_n\left(\frac{e^{4\sigma}}{r_c^2}\partial_\phi (e^{-4\sigma}\partial_\phi f_n)-M_\Phi^2 f_n\right)S_n
	\nonumber \\
	&-\frac{\lambda_S^{(5)}}{r_c}\sum_n f_nS_n\sum_{m l} f_mf_l S_mS_l.
\end{align}
The 4D EoM for a scalar of mode $\hat{n}$ can be projected out   by multiplying a factor of $e^{-4\sigma}f_{\hat{n}}$
on both sides of eq.~\eqref{5DEoM}  and integrating out the fifth dimension.  Using the orthonormal condition, eq.~\eqref{snorm}, we can see that,
eq.~\eqref{5DEoM} reduces to
	\begin{align}\label{4DSEoM}
	0=	\partial^2 S_{\hat{n}}+m_{\hat{n}}^2 S_{\hat{n}}-\sum_{nml}  \lambda_{S,\hat{n}nml} S_n S_m S_l,
\end{align}
where the effective 4D mass $m_{{\hat{n}}}$ is identified as
\begin{align}\label{mid}
m_{\hat{n}}^2&\equiv - \int_{-\pi}^{\pi}d\phi f_{\hat{n}}\left(\frac{1}{r_c^2}\partial_\phi(e^{-4\sigma}\partial_\phi f_n)-e^{-4\sigma}M_\Phi^2f_n\right)
\nonumber \\
&	= \int_{-\pi}^{\pi}d\phi f_{\hat{n}}\left( e^{-2\sigma}m_{n}^2f_n\right),
\end{align}
with the orthonormal condition applied to the last step, and the quartic coupling is defined by
\begin{align}\label{quartic}
	\lambda_{S,\hat{n}nml}\equiv \frac{\lambda_S^{(5)}}{r_c}\int_{-\pi}^{\pi}d\phi e^{-4\sigma}f_{\hat{n}} f_n f_m  f_l.
\end{align}
 Eq.~\eqref{mid} is nothing but the EoM given by eq.~\eqref{sEoM}. Therefore, even after including the self-interaction in  the 5D EoM, the 4D mass identification is no affected, and the previous EoM for obtaining the profile is valid. Moreover, eq.~\eqref{lambdaS} can be readily reproduced via eq.~\eqref{quartic}.

 It should be mentioned that, the derivation of  eq.~\eqref{sEoM} follows the philosophy of KK decomposition~\cite{Ponton:2012bi}, which is defined in the free action, while interactions are treated perturbatively after inserting the KK decomposition~\cite{Ponton:2012bi}. To validate this perturbative treatment, we apply the higher-mode profiles~\cite{Gherghetta:2000qt} and find that, for the zero-mode  EoM concerned here, the quartic couplings associated with $S_0$ are well within the perturbative region. In particular, the couplings of $S_0 S_iS_jS_k$, $S_0^2 S_iS_j$ and $S_0^3 S_i$ for low-lying $i,j,k>0$ are many orders-of-magnitude smaller  than  $\lambda_{S,0000}\equiv\lambda_{S,0}$, which will be shown later to be $\lambda_{S,0}\lesssim\mathcal	{O}(0.1)$.

In the following, we proceed with the    DM evolution by a simple two-step   freeze-in production under some proper assumptions, and show that   eqs.~\eqref{lambdaP} and~\eqref{lambdaS} can successfully predict the large parameter span in terms of  the $\mathcal{O}(1)$ 5D parameters: $a,\xi,\lambda_{P}^{(5)}$.

\TT{Uniting freeze-in self-interacting DM:} For simplicity, we  consider the case where the KK scalars $S_n$ have an unbroken $Z_2$ symmetry in the 4D regime such that  the   vacuum expectation values vanish  $\langle S_n\rangle=0$.  In this regime,  there is no mass mixing between the zero mode $S_0$ and the physical Higgs boson $h$.
Being the lightest KK scalar, $S_0$ can be a stable DM candidate, with  the mass  set by the SM Higgs vacuum
\begin{align}\label{DMmass}
	m_{S_0}=\sqrt{\lambda_{P,0}}v,
\end{align}
which would be fixed once the portal coupling is pinned down by fitting the observed   relic abundance  from freeze-in production. To wit, the mass and relic density of  the classically scale-invariant DM share a common origin from  a portal coupling~\cite{McDonald:2001vt,Kang:2015aqa}.

It should be mentioned that,
in explicit DM models  with an  implicitly assumed  initial reheating mechanism, the production and evolution history of the scalar DM can be more involved than the conventional freeze-in pattern~\cite{Chu:2011be,Bernal:2015ova,Heeba:2018wtf}.  Generically,
at the gauge-symmetric phase,   the DM   interacts with the Higgs doublet $H$ via  $2\to 2$ scattering $HH\to S_0S_0$. At the gauge-broken phase, the dominant production comes from the Higgs boson     decay  $h\to S_0 S_0$.
 As  temperature goes down to the Higgs boson mass $T\lesssim m_h$,  the  $h$-particle density    becomes Boltzmann suppressed  and the DM finally freezes in. Note that, if the DM  has relatively strong self-interactions,  the   particle  density may get modified via particle-changing  $2\leftrightharpoons4$ processes~\cite{Bernal:2015xba}, and the DM could thermalize itself with a different temperature $T^\prime$ and  further undergo the \textit{dark freeze-out}~\cite{Feng:2008mu,Ackerman:mha}. However, the \textit{dark freeze-out} process depends on the  initial temperature ratio of the dark sector and the visible SM bath, $\chi \equiv T/T^\prime$, and it could be absent if  the initial ratio  is  too large ($\chi \gg 1$) to establish DM self-thermalization, which may originate from  an explicit reheating dynamics~\cite{Hodges:1993yb,Berezhiani:1995am,Feng:2008mu}.

On the other hand,  additional production channels of the DM $S_0$ are also possible from the portal interaction  $S_0 S_i H^\dagger H$, with $i>0$. For the low-lying   KK modes, their masses are given by~\cite{Gherghetta:2000qt}
\begin{align}
	m_{S_i}
	\simeq \left(i+\frac{\sqrt{a+4}}{2}-\frac{3}{4}\right)k \pi e^{-k r_c\pi}\gtrsim 2~\text{TeV},
\end{align}
and  the  portal coupling  between $S_0$ and $S_i$  turns out to be
\begin{align}
	\lambda_{P,0i}=\lambda_{P}^{(5)}e^{-2\sigma(\pi)}f_0(\pi) f_i(\pi) \simeq \mathcal{O}(10^{-5}).
\end{align}
In the early Universe, the productions  via $HH\to S_0 S_i$ before EWSB and  $S_i\to h S_0$ after EWSB could  contribute to the final $S_0$ relic density  given that the portal coupling $\lambda_{P,0i}$  is sufficiently larger than   $\lambda_{P,0}\simeq \mathcal{O}(10^{-10})$.  Nevertheless, the influence may be   small or even negligible,     if the  initial dynamics predicts such a quite low  reheating temperature $T_{reh}$  that the thermal mass $M_{H}(T)$  cannot significantly exceed  $m_{S_i}/2$. In this case,   the $HH\to S_0 S_i$   channel  is   endothermic and thus suppressed, or the production of heavier KK scalars   could even not occur. In some complicated cases, on the other hand,  the heavier KK scalars may decay or annihilate  away quickly through gravitational effects~\cite{Bernal:2020fvw}, thus produce negligible $S_0$ yield.

Given the above uncertainties, we will   assume in the following for simplicity that  the \textit{dark freeze-out} process is absent  and  the  DM relic abundance is fixed  by   the two-step production in the early Universe:  $H H\to S_0S_0$ before EWSB  and $h\to S_0S_0$ after EWSB.  The Boltzmann equations for the two  processes can then be written as
\begin{align}
	\frac{dY_1}{dT}=-\frac{2\gamma_{HH\to S_0S_0}}{s_{\rm SM}\mathcal{H} T},\qquad  \frac{dY_2}{dT}=-\frac{2\gamma_{h\to S_0S_0}}{s_{\rm SM} \mathcal{H} T},
\end{align}
where the factor 2 accounts for the fact that DM is pair-produced. The Hubble rate is given by $\mathcal{H}(T)=1.66 \sqrt{g_*^\rho}T^2/M_{Pl}$  and the entropy density of the SM plasma  $s_{\rm SM}=2\pi^2 g_*^sT^3/45$, with the relativistic degrees of freedom $g_{*}^s\approx g_{*}^\rho\approx 106.75$ during  the production epoch. We use $Y_{1,2}$ to denote the  DM yields produced before and after the EWSB. The collision terms $\gamma_{HH\to S_0S_0}$  and  $\gamma_{h\to S_0S_0}$ are determined to be
\begin{align}
	\gamma_{HH\to S_0S_0}&=\frac{2\pi^2 T}{(2\pi)^6}\int_{4M_H^2}^\infty d\hat{s} \sqrt{\hat{s}} (\hat{s}-4M_H^2)K_1(\sqrt{\hat{s}}/T)
	\nonumber \\
	&\times \sigma_{HH\to S_0S_0},
	\\[0.2mm]
	\gamma_{h\to S_0S_0}&=\frac{m_h^2T }{2\pi^2}\Gamma_{h\to S_0S_0} K_1(m_h/T),
\end{align}
respectively. Here $K_1$ is the first modified Bessel function of the second kind. $m_h\simeq 125$~GeV is the  vacuum mass of the SM Higgs boson and the thermal mass $M_H$ at $T>T_c$ is given by~\cite{Weldon:1982bn,Quiros:1999jp}
\begin{align}
	M_H^2(T)= \left(\frac{3}{16}g_2^2+\frac{1}{16}g_1^2+\frac{1}{4}y_t^2+\frac{1}{8}\lambda_H\right)(T^2-T_c^2),
\end{align}
where $g_{2}$~($g_{1}$) is the $SU(2)_L$~($U(1)_Y$) gauge coupling and $y_t$ the top-quark Yukawa coupling.  $T_c\simeq 164$~GeV denotes the critical temperature of electroweak phase transition.
The cross section $\sigma_{HH\to S_0S_0}$ and decay rate $\Gamma_{h\to S_0S_0}$ are given by
\begin{align}\label{rates}
	\sigma_{HH\to S_0S_0}&=\frac{g_H}{8\pi}\frac{\lambda_{P,0}^2}{\sqrt{\hat{s}(\hat{s}-4M_H^2)}},	
	\\
	\Gamma_{h\to S_0S_0}&=\frac{\lambda_{P,0}^2}{8\pi}\frac{v^2}{m_h},
\end{align}
where $g_H=2$ accounts for the two gauge components of the Higgs doublet,  and we have neglected the scalar DM mass since $m_{S_0}$ is expected to be at MeV scale.

Integrating $Y_1$ from $T=T_c$ to $\infty$ gives the yield produced before EWSB, and  integrating $Y_2$ from $T=0$ to $T_c$ gives the yield produced by the Higgs decay after EWSB. Note that, besides the subdominant contribution from scattering processes after EWSB, we have also  neglected a small yield  produced from a short period of time  around  $T\approx T_c$~\cite{Heeba:2018wtf}.  The total yield is then found to be
\begin{align}
	Y_f=Y_1+Y_2\approx\left(1.28\times 10^{10}+ 8.58 \times 10^{12}\right)\lambda_{P,0}^2.
\end{align}
It can be seen that, the primary production comes from the Higgs decay   after EWSB.  Using the current values of entropy ($s_{{\rm SM},0}$) and critical energy ($\rho_c$) densities~\cite{Zyla:2020zbs}, we can estimate  the DM relic abundance  via
\begin{align}
	\Omega_{S_0} h^2=\frac{m_{S_0} Y_f s_{{\rm SM},0}}{\rho_{c}/h^2}
	\approx2.74\times 10^{5} \left(\frac{m_{S_0}}{\rm MeV}\right)Y_f.
\end{align}
\begin{figure}[t]
	\centering	
	\includegraphics[scale=0.55]{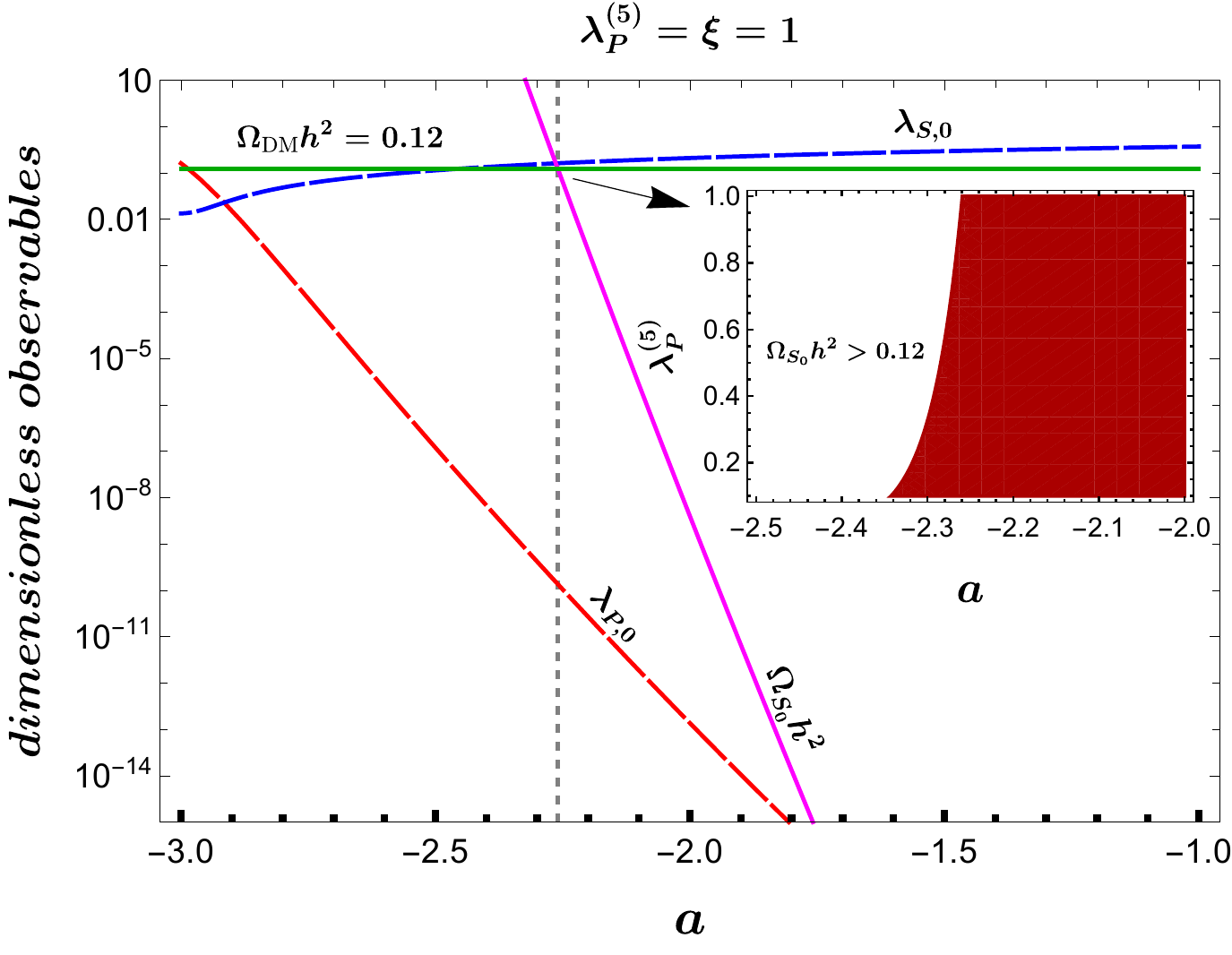}
	\caption{\label{sprofile-plot} $a$-dependence of dimensionless observables.  The cross point of $\Omega_{S_0}h^2= \Omega_{\text{DM}} h^2=0.12$ predicts  $a\approx -2.26$ (the vertical line) with a fixed value $\lambda_{P}^{(5)}=1$, while the general case with varying $\lambda_{P}^{(5)}$ is also shown in the subfigure. }
	
\end{figure}

It becomes clear that,   the DM mass $m_{S_0}$, relic density $\Omega_{S_0} h^2$ and the Higgs portal coupling $\lambda_{P,0}$   can all be well  estimated  by a single $\mathcal{O}(1)$ $a$  since  their    exponential behavior is controlled by $a$.  We plot this feature in  figure~\ref{sprofile-plot} by setting $\lambda_P^{(5)}=1$ to  show the $a$-dependence of the portal coupling $\lambda_{P,0}$ (red)  and  relic abundance $\Omega_{S_0} h^2$  (magenta). Besides, by varying $\lambda_P^{(5)}$, we further  show in the subfigure the correlation between $\lambda_{P}^{(5)}$ and $a$ in generating the DM relic density, where the dark-red region denotes $\Omega_{S_0} h^2\leqslant 0.12$. It can be seen that,  the fitting of $\Omega_{\text{DM}} h^2=0.12$~\cite{Aghanim:2018eyx} exhibits a  weak dependence on the parameter $\lambda_{P}^{(5)}$. This property   allows us to obtain  a good estimate of the parameter $a$ for  varying $\lambda_{P}^{(5)}$ in the natural regime $\mathcal{O}(0.1-1)$.   As a reference point, we set  $\lambda_P^{(5)}=1$ to obtain $a\approx -2.26$, as shown by the vertical line in figure~\ref{sprofile-plot}. Then using the estimated value of $a$, we can predict  the portal coupling and mass  to be
\begin{align}
	\lambda_{P,0}= 1.34\times 10^{-10}\lambda_{P}^{(5)},~m_{S_0}=2.85 \sqrt{\lambda_{P}^{(5)}}~{\rm MeV}.
\end{align}

\begin{figure}[t]
	\centering	
	\includegraphics[scale=0.55]{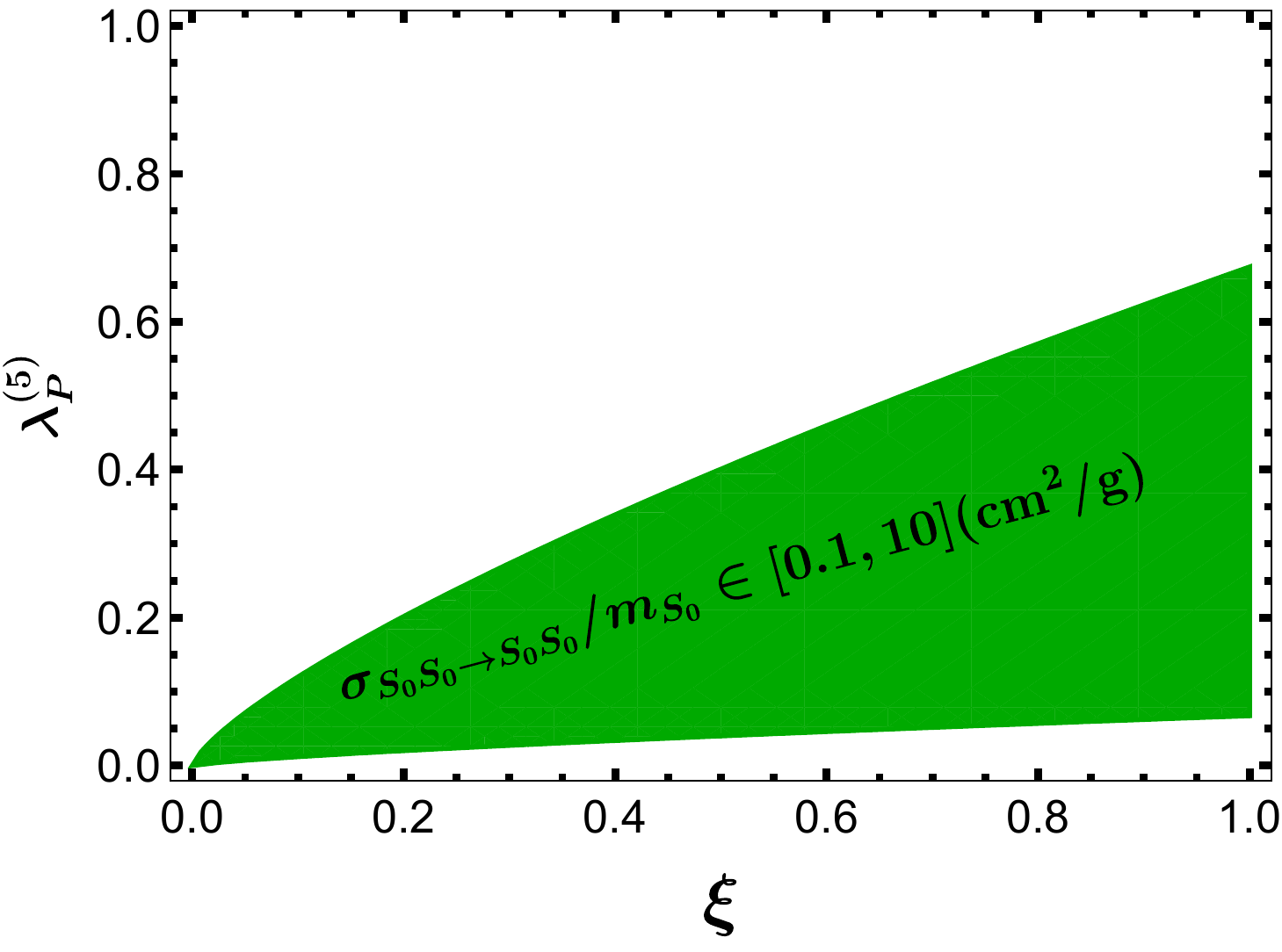}
	\caption{\label{sigma-m} DM self-interacting cross section over mass, with a fixed value $a\approx -2.26$ obtained from fitting   the DM relic density. The dark-green region corresponds to a ratio $\sigma_{S_0S_0\to S_0S_0}/m_{S_0}$ in the range: $0.1- 10~\text{cm}^2/\text{g}$.}
	
\end{figure}

In addition, we  also show the $a$-dependence of the self-interacting coupling $\lambda_{S,0}$ (blue) by fixing $\xi=1$. It can be readily seen that $\lambda_{S,0}$ is rather stable in terms of varying $a$, which can also be inferred from eq.~\eqref{lambdaS}.  This property implies that, we can  fix $a\approx -2.26 $ obtained from fitting the relic density to exploit the DM self-interaction in light of the free parameters $\lambda_{P}^{(5)},\xi$. In this context, a similar order  with respect to $\lambda_H=2m_h^2/v^2$ is induced:
$\lambda_{S,0}\approx 0.16\xi$.
Remarkably,  it predicts a  ratio of   $S_0S_0\to S_0S_0$ cross section over mass $m_{S_0}$ as
\begin{align}
	\frac{\sigma_{S_0S_0\to S_0S_0}}{m_{S_0}}\approx  \frac{22\; \xi^2}{(\lambda_{P}^{(5)})^{3/2}}~(\text{cm}^2/\text{g}).
\end{align}
We show in figure~\ref{sigma-m} the region of $\sigma_{S_0S_0\to S_0S_0}/m_{S_0}$ that fits the  favored range $0.1- 10~\text{cm}^2/\text{g}$,  validating the naturalness criterion on these fundamental parameters.  Note here that,   the   range is simply taken as a  moderate estimate  to infer the ability of solving the small-scale issues, such as the \textit{too-big-to-fail} and  \textit{missing satellites} problems. A detailed simulation for these explanations requires taking various uncertainties from other observational constraints. For more details, one can consult from the recent review~\cite{Tulin:2017ara}.  In short, we have demonstrated that,   the  classically scale-invariant scalar DM from the WED can simultaneously exhibit an exponentially suppressed Higgs portal to implement the freeze-in production, and  a self-interacting pattern: $\sigma_{S_0S_0\to S_0S_0}/m_{S_0} \in [0.1- 10]~\text{cm}^2/\text{g}$ to alleviate the small-scale problems~\cite{Spergel:1999mh}   from 5D \textit{natural} couplings.

\TT{Summary:} We have presented an embedding of a classically scale-invariant DM singlet in the warped extra dimension.  With an exponentially suppressed profile in the extra dimension, a feeble portal coupling at $\mathcal{O}(10^{-10})$ and a relatively large quartic coupling at $\mathcal{O}(0.1)$ can be naturally generated by well-controlled fundamental parameters. Under some proper initial assumptions, the DM relic abundance can be approximately generated by a two-step freeze-in process.  In this regime, we have shown that, it successfully explains the numerical coincidence for freeze-in self-interacting DM,   which on one hand fits the current relic density, and on the other hand,     alleviate the small-scale problems   via considerable self-interactions.

We thank Xin-Shuai Yan for useful discussions.
This work is  supported by the National Natural Science Foundation of China under Grant
No.11947131 as well as by the CCNU-QLPL Innovation Fund (QLPL2019P01). S. L.  is also supported by the Fundamental Research Funds for the Central Universities under Grant $\text{No}.~\text{2020YBZZ074}$.

\bibliographystyle{apsrev4-1}
\bibliography{reference}

\begin{thebibliography}{32}%
\makeatletter
\providecommand \@ifxundefined [1]{%
 \@ifx{#1\undefined}
}%
\providecommand \@ifnum [1]{%
 \ifnum #1\expandafter \@firstoftwo
 \else \expandafter \@secondoftwo
 \fi
}%
\providecommand \@ifx [1]{%
 \ifx #1\expandafter \@firstoftwo
 \else \expandafter \@secondoftwo
 \fi
}%
\providecommand \natexlab [1]{#1}%
\providecommand \enquote  [1]{``#1''}%
\providecommand \bibnamefont  [1]{#1}%
\providecommand \bibfnamefont [1]{#1}%
\providecommand \citenamefont [1]{#1}%
\providecommand \href@noop [0]{\@secondoftwo}%
\providecommand \href [0]{\begingroup \@sanitize@url \@href}%
\providecommand \@href[1]{\@@startlink{#1}\@@href}%
\providecommand \@@href[1]{\endgroup#1\@@endlink}%
\providecommand \@sanitize@url [0]{\catcode `\\12\catcode `\$12\catcode
  `\&12\catcode `\#12\catcode `\^12\catcode `\_12\catcode `\%12\relax}%
\providecommand \@@startlink[1]{}%
\providecommand \@@endlink[0]{}%
\providecommand \url  [0]{\begingroup\@sanitize@url \@url }%
\providecommand \@url [1]{\endgroup\@href {#1}{\urlprefix }}%
\providecommand \urlprefix  [0]{URL }%
\providecommand \Eprint [0]{\href }%
\providecommand \doibase [0]{http://dx.doi.org/}%
\providecommand \selectlanguage [0]{\@gobble}%
\providecommand \bibinfo  [0]{\@secondoftwo}%
\providecommand \bibfield  [0]{\@secondoftwo}%
\providecommand \translation [1]{[#1]}%
\providecommand \BibitemOpen [0]{}%
\providecommand \bibitemStop [0]{}%
\providecommand \bibitemNoStop [0]{.\EOS\space}%
\providecommand \EOS [0]{\spacefactor3000\relax}%
\providecommand \BibitemShut  [1]{\csname bibitem#1\endcsname}%
\let\auto@bib@innerbib\@empty
\bibitem [{\citenamefont {Randall}\ and\ \citenamefont
  {Sundrum}(1999)}]{Randall:1999ee}%
  \BibitemOpen
  \bibfield  {author} {\bibinfo {author} {\bibfnamefont {L.}~\bibnamefont
  {Randall}}\ and\ \bibinfo {author} {\bibfnamefont {R.}~\bibnamefont
  {Sundrum}},\ }\href {\doibase 10.1103/PhysRevLett.83.3370} {\bibfield
  {journal} {\bibinfo  {journal} {Phys. Rev. Lett.}\ }\textbf {\bibinfo
  {volume} {83}},\ \bibinfo {pages} {3370} (\bibinfo {year} {1999})},\ \Eprint
  {http://arxiv.org/abs/hep-ph/9905221} {arXiv:hep-ph/9905221} \BibitemShut
  {NoStop}%
\bibitem [{\citenamefont {Grossman}\ and\ \citenamefont
  {Neubert}(2000)}]{Grossman:1999ra}%
  \BibitemOpen
  \bibfield  {author} {\bibinfo {author} {\bibfnamefont {Y.}~\bibnamefont
  {Grossman}}\ and\ \bibinfo {author} {\bibfnamefont {M.}~\bibnamefont
  {Neubert}},\ }\href {\doibase 10.1016/S0370-2693(00)00054-X} {\bibfield
  {journal} {\bibinfo  {journal} {Phys. Lett. B}\ }\textbf {\bibinfo {volume}
  {474}},\ \bibinfo {pages} {361} (\bibinfo {year} {2000})},\ \Eprint
  {http://arxiv.org/abs/hep-ph/9912408} {arXiv:hep-ph/9912408} \BibitemShut
  {NoStop}%
\bibitem [{\citenamefont {Huber}\ and\ \citenamefont
  {Shafi}(2001)}]{Huber:2000ie}%
  \BibitemOpen
  \bibfield  {author} {\bibinfo {author} {\bibfnamefont {S.~J.}\ \bibnamefont
  {Huber}}\ and\ \bibinfo {author} {\bibfnamefont {Q.}~\bibnamefont {Shafi}},\
  }\href {\doibase 10.1016/S0370-2693(00)01399-X} {\bibfield  {journal}
  {\bibinfo  {journal} {Phys. Lett. B}\ }\textbf {\bibinfo {volume} {498}},\
  \bibinfo {pages} {256} (\bibinfo {year} {2001})},\ \Eprint
  {http://arxiv.org/abs/hep-ph/0010195} {arXiv:hep-ph/0010195} \BibitemShut
  {NoStop}%
\bibitem [{\citenamefont {Gherghetta}\ and\ \citenamefont
  {Pomarol}(2000)}]{Gherghetta:2000qt}%
  \BibitemOpen
  \bibfield  {author} {\bibinfo {author} {\bibfnamefont {T.}~\bibnamefont
  {Gherghetta}}\ and\ \bibinfo {author} {\bibfnamefont {A.}~\bibnamefont
  {Pomarol}},\ }\href {\doibase 10.1016/S0550-3213(00)00392-8} {\bibfield
  {journal} {\bibinfo  {journal} {Nucl. Phys. B}\ }\textbf {\bibinfo {volume}
  {586}},\ \bibinfo {pages} {141} (\bibinfo {year} {2000})},\ \Eprint
  {http://arxiv.org/abs/hep-ph/0003129} {arXiv:hep-ph/0003129} \BibitemShut
  {NoStop}%
\bibitem [{\citenamefont {Hall}\ \emph {et~al.}(2010)\citenamefont {Hall},
  \citenamefont {Jedamzik}, \citenamefont {March-Russell},\ and\ \citenamefont
  {West}}]{Hall:2009bx}%
  \BibitemOpen
  \bibfield  {author} {\bibinfo {author} {\bibfnamefont {L.~J.}\ \bibnamefont
  {Hall}}, \bibinfo {author} {\bibfnamefont {K.}~\bibnamefont {Jedamzik}},
  \bibinfo {author} {\bibfnamefont {J.}~\bibnamefont {March-Russell}}, \ and\
  \bibinfo {author} {\bibfnamefont {S.~M.}\ \bibnamefont {West}},\ }\href
  {\doibase 10.1007/JHEP03(2010)080} {\bibfield  {journal} {\bibinfo  {journal}
  {JHEP}\ }\textbf {\bibinfo {volume} {03}},\ \bibinfo {pages} {080} (\bibinfo
  {year} {2010})},\ \Eprint {http://arxiv.org/abs/0911.1120} {arXiv:0911.1120
  [hep-ph]} \BibitemShut {NoStop}%
\bibitem [{\citenamefont {Yaguna}(2011)}]{Yaguna:2011qn}%
  \BibitemOpen
  \bibfield  {author} {\bibinfo {author} {\bibfnamefont {C.~E.}\ \bibnamefont
  {Yaguna}},\ }\href {\doibase 10.1007/JHEP08(2011)060} {\bibfield  {journal}
  {\bibinfo  {journal} {JHEP}\ }\textbf {\bibinfo {volume} {08}},\ \bibinfo
  {pages} {060} (\bibinfo {year} {2011})},\ \Eprint
  {http://arxiv.org/abs/1105.1654} {arXiv:1105.1654 [hep-ph]} \BibitemShut
  {NoStop}%
\bibitem [{\citenamefont {Blennow}\ \emph {et~al.}(2014)\citenamefont
  {Blennow}, \citenamefont {Fernandez-Martinez},\ and\ \citenamefont
  {Zaldivar}}]{Blennow:2013jba}%
  \BibitemOpen
  \bibfield  {author} {\bibinfo {author} {\bibfnamefont {M.}~\bibnamefont
  {Blennow}}, \bibinfo {author} {\bibfnamefont {E.}~\bibnamefont
  {Fernandez-Martinez}}, \ and\ \bibinfo {author} {\bibfnamefont
  {B.}~\bibnamefont {Zaldivar}},\ }\href {\doibase
  10.1088/1475-7516/2014/01/003} {\bibfield  {journal} {\bibinfo  {journal}
  {JCAP}\ }\textbf {\bibinfo {volume} {01}},\ \bibinfo {pages} {003} (\bibinfo
  {year} {2014})},\ \Eprint {http://arxiv.org/abs/1309.7348} {arXiv:1309.7348
  [hep-ph]} \BibitemShut {NoStop}%
\bibitem [{\citenamefont {Heikinheimo}\ \emph {et~al.}(2016)\citenamefont
  {Heikinheimo}, \citenamefont {Tenkanen}, \citenamefont {Tuominen},\ and\
  \citenamefont {Vaskonen}}]{Heikinheimo:2016yds}%
  \BibitemOpen
  \bibfield  {author} {\bibinfo {author} {\bibfnamefont {M.}~\bibnamefont
  {Heikinheimo}}, \bibinfo {author} {\bibfnamefont {T.}~\bibnamefont
  {Tenkanen}}, \bibinfo {author} {\bibfnamefont {K.}~\bibnamefont {Tuominen}},
  \ and\ \bibinfo {author} {\bibfnamefont {V.}~\bibnamefont {Vaskonen}},\
  }\href {\doibase 10.1103/PhysRevD.94.063506} {\bibfield  {journal} {\bibinfo
  {journal} {Phys. Rev. D}\ }\textbf {\bibinfo {volume} {94}},\ \bibinfo
  {pages} {063506} (\bibinfo {year} {2016})},\ \bibinfo {note} {[Erratum:
  Phys.Rev.D 96, 109902 (2017)]},\ \Eprint {http://arxiv.org/abs/1604.02401}
  {arXiv:1604.02401 [astro-ph.CO]} \BibitemShut {NoStop}%
\bibitem [{\citenamefont {Heeba}\ \emph {et~al.}(2018)\citenamefont {Heeba},
  \citenamefont {Kahlhoefer},\ and\ \citenamefont {St\"ocker}}]{Heeba:2018wtf}%
  \BibitemOpen
  \bibfield  {author} {\bibinfo {author} {\bibfnamefont {S.}~\bibnamefont
  {Heeba}}, \bibinfo {author} {\bibfnamefont {F.}~\bibnamefont {Kahlhoefer}}, \
  and\ \bibinfo {author} {\bibfnamefont {P.}~\bibnamefont {St\"ocker}},\ }\href
  {\doibase 10.1088/1475-7516/2018/11/048} {\bibfield  {journal} {\bibinfo
  {journal} {JCAP}\ }\textbf {\bibinfo {volume} {11}},\ \bibinfo {pages} {048}
  (\bibinfo {year} {2018})},\ \Eprint {http://arxiv.org/abs/1809.04849}
  {arXiv:1809.04849 [hep-ph]} \BibitemShut {NoStop}%
\bibitem [{\citenamefont {Elahi}\ \emph {et~al.}(2015)\citenamefont {Elahi},
  \citenamefont {Kolda},\ and\ \citenamefont {Unwin}}]{Elahi:2014fsa}%
  \BibitemOpen
  \bibfield  {author} {\bibinfo {author} {\bibfnamefont {F.}~\bibnamefont
  {Elahi}}, \bibinfo {author} {\bibfnamefont {C.}~\bibnamefont {Kolda}}, \ and\
  \bibinfo {author} {\bibfnamefont {J.}~\bibnamefont {Unwin}},\ }\href
  {\doibase 10.1007/JHEP03(2015)048} {\bibfield  {journal} {\bibinfo  {journal}
  {JHEP}\ }\textbf {\bibinfo {volume} {03}},\ \bibinfo {pages} {048} (\bibinfo
  {year} {2015})},\ \Eprint {http://arxiv.org/abs/1410.6157} {arXiv:1410.6157
  [hep-ph]} \BibitemShut {NoStop}%
\bibitem [{\citenamefont {Kim}\ and\ \citenamefont
  {McDonald}(2018)}]{Kim:2017mtc}%
  \BibitemOpen
  \bibfield  {author} {\bibinfo {author} {\bibfnamefont {J.}~\bibnamefont
  {Kim}}\ and\ \bibinfo {author} {\bibfnamefont {J.}~\bibnamefont {McDonald}},\
  }\href {\doibase 10.1103/PhysRevD.98.023533} {\bibfield  {journal} {\bibinfo
  {journal} {Phys. Rev. D}\ }\textbf {\bibinfo {volume} {98}},\ \bibinfo
  {pages} {023533} (\bibinfo {year} {2018})},\ \Eprint
  {http://arxiv.org/abs/1709.04105} {arXiv:1709.04105 [hep-ph]} \BibitemShut
  {NoStop}%
\bibitem [{\citenamefont {Kim}\ and\ \citenamefont
  {Mcdonald}(2018)}]{Kim:2018xsp}%
  \BibitemOpen
  \bibfield  {author} {\bibinfo {author} {\bibfnamefont {J.}~\bibnamefont
  {Kim}}\ and\ \bibinfo {author} {\bibfnamefont {J.}~\bibnamefont {Mcdonald}},\
  }\href {\doibase 10.1103/PhysRevD.98.123503} {\bibfield  {journal} {\bibinfo
  {journal} {Phys. Rev. D}\ }\textbf {\bibinfo {volume} {98}},\ \bibinfo
  {pages} {123503} (\bibinfo {year} {2018})},\ \Eprint
  {http://arxiv.org/abs/1804.02661} {arXiv:1804.02661 [hep-ph]} \BibitemShut
  {NoStop}%
\bibitem [{\citenamefont {Borah}\ \emph {et~al.}(2018)\citenamefont {Borah},
  \citenamefont {Karmakar},\ and\ \citenamefont {Nanda}}]{Borah:2018gjk}%
  \BibitemOpen
  \bibfield  {author} {\bibinfo {author} {\bibfnamefont {D.}~\bibnamefont
  {Borah}}, \bibinfo {author} {\bibfnamefont {B.}~\bibnamefont {Karmakar}}, \
  and\ \bibinfo {author} {\bibfnamefont {D.}~\bibnamefont {Nanda}},\ }\href
  {\doibase 10.1088/1475-7516/2018/07/039} {\bibfield  {journal} {\bibinfo
  {journal} {JCAP}\ }\textbf {\bibinfo {volume} {07}},\ \bibinfo {pages} {039}
  (\bibinfo {year} {2018})},\ \Eprint {http://arxiv.org/abs/1805.11115}
  {arXiv:1805.11115 [hep-ph]} \BibitemShut {NoStop}%
\bibitem [{\citenamefont {Goudelis}\ \emph {et~al.}(2018)\citenamefont
  {Goudelis}, \citenamefont {Mohan},\ and\ \citenamefont
  {Sengupta}}]{Goudelis:2018xqi}%
  \BibitemOpen
  \bibfield  {author} {\bibinfo {author} {\bibfnamefont {A.}~\bibnamefont
  {Goudelis}}, \bibinfo {author} {\bibfnamefont {K.~A.}\ \bibnamefont {Mohan}},
  \ and\ \bibinfo {author} {\bibfnamefont {D.}~\bibnamefont {Sengupta}},\
  }\href {\doibase 10.1007/JHEP10(2018)014} {\bibfield  {journal} {\bibinfo
  {journal} {JHEP}\ }\textbf {\bibinfo {volume} {10}},\ \bibinfo {pages} {014}
  (\bibinfo {year} {2018})},\ \Eprint {http://arxiv.org/abs/1807.06642}
  {arXiv:1807.06642 [hep-ph]} \BibitemShut {NoStop}%
\bibitem [{\citenamefont {McDonald}(2002)}]{McDonald:2001vt}%
  \BibitemOpen
  \bibfield  {author} {\bibinfo {author} {\bibfnamefont {J.}~\bibnamefont
  {McDonald}},\ }\href {\doibase 10.1103/PhysRevLett.88.091304} {\bibfield
  {journal} {\bibinfo  {journal} {Phys. Rev. Lett.}\ }\textbf {\bibinfo
  {volume} {88}},\ \bibinfo {pages} {091304} (\bibinfo {year} {2002})},\
  \Eprint {http://arxiv.org/abs/hep-ph/0106249} {arXiv:hep-ph/0106249}
  \BibitemShut {NoStop}%
\bibitem [{\citenamefont {Kang}(2015)}]{Kang:2015aqa}%
  \BibitemOpen
  \bibfield  {author} {\bibinfo {author} {\bibfnamefont {Z.}~\bibnamefont
  {Kang}},\ }\href {\doibase 10.1016/j.physletb.2015.10.031} {\bibfield
  {journal} {\bibinfo  {journal} {Phys. Lett. B}\ }\textbf {\bibinfo {volume}
  {751}},\ \bibinfo {pages} {201} (\bibinfo {year} {2015})},\ \Eprint
  {http://arxiv.org/abs/1505.06554} {arXiv:1505.06554 [hep-ph]} \BibitemShut
  {NoStop}%
\bibitem [{\citenamefont {Spergel}\ and\ \citenamefont
  {Steinhardt}(2000)}]{Spergel:1999mh}%
  \BibitemOpen
  \bibfield  {author} {\bibinfo {author} {\bibfnamefont {D.~N.}\ \bibnamefont
  {Spergel}}\ and\ \bibinfo {author} {\bibfnamefont {P.~J.}\ \bibnamefont
  {Steinhardt}},\ }\href {\doibase 10.1103/PhysRevLett.84.3760} {\bibfield
  {journal} {\bibinfo  {journal} {Phys. Rev. Lett.}\ }\textbf {\bibinfo
  {volume} {84}},\ \bibinfo {pages} {3760} (\bibinfo {year} {2000})},\ \Eprint
  {http://arxiv.org/abs/astro-ph/9909386} {arXiv:astro-ph/9909386} \BibitemShut
  {NoStop}%
\bibitem [{\citenamefont {Tulin}\ and\ \citenamefont
  {Yu}(2018)}]{Tulin:2017ara}%
  \BibitemOpen
  \bibfield  {author} {\bibinfo {author} {\bibfnamefont {S.}~\bibnamefont
  {Tulin}}\ and\ \bibinfo {author} {\bibfnamefont {H.-B.}\ \bibnamefont {Yu}},\
  }\href {\doibase 10.1016/j.physrep.2017.11.004} {\bibfield  {journal}
  {\bibinfo  {journal} {Phys. Rept.}\ }\textbf {\bibinfo {volume} {730}},\
  \bibinfo {pages} {1} (\bibinfo {year} {2018})},\ \Eprint
  {http://arxiv.org/abs/1705.02358} {arXiv:1705.02358 [hep-ph]} \BibitemShut
  {NoStop}%
\bibitem [{\citenamefont {Goldberger}\ and\ \citenamefont
  {Wise}(1999)}]{Goldberger:1999wh}%
  \BibitemOpen
  \bibfield  {author} {\bibinfo {author} {\bibfnamefont {W.~D.}\ \bibnamefont
  {Goldberger}}\ and\ \bibinfo {author} {\bibfnamefont {M.~B.}\ \bibnamefont
  {Wise}},\ }\href {\doibase 10.1103/PhysRevD.60.107505} {\bibfield  {journal}
  {\bibinfo  {journal} {Phys. Rev. D}\ }\textbf {\bibinfo {volume} {60}},\
  \bibinfo {pages} {107505} (\bibinfo {year} {1999})},\ \Eprint
  {http://arxiv.org/abs/hep-ph/9907218} {arXiv:hep-ph/9907218} \BibitemShut
  {NoStop}%
\bibitem [{\citenamefont {Ponton}(2012)}]{Ponton:2012bi}%
  \BibitemOpen
  \bibfield  {author} {\bibinfo {author} {\bibfnamefont {E.}~\bibnamefont
  {Ponton}},\ }in\ \href {\doibase 10.1142/9789814390163_0007} {\emph {\bibinfo
  {booktitle} {{Theoretical Advanced Study Institute in Elementary Particle
  Physics}: {The Dark Secrets of the Terascale}}}}\ (\bibinfo {year} {2012})\
  \Eprint {http://arxiv.org/abs/1207.3827} {arXiv:1207.3827 [hep-ph]}
  \BibitemShut {NoStop}%
\bibitem [{\citenamefont {Chu}\ \emph {et~al.}(2012)\citenamefont {Chu},
  \citenamefont {Hambye},\ and\ \citenamefont {Tytgat}}]{Chu:2011be}%
  \BibitemOpen
  \bibfield  {author} {\bibinfo {author} {\bibfnamefont {X.}~\bibnamefont
  {Chu}}, \bibinfo {author} {\bibfnamefont {T.}~\bibnamefont {Hambye}}, \ and\
  \bibinfo {author} {\bibfnamefont {M.~H.}\ \bibnamefont {Tytgat}},\ }\href
  {\doibase 10.1088/1475-7516/2012/05/034} {\bibfield  {journal} {\bibinfo
  {journal} {JCAP}\ }\textbf {\bibinfo {volume} {05}},\ \bibinfo {pages} {034}
  (\bibinfo {year} {2012})},\ \Eprint {http://arxiv.org/abs/1112.0493}
  {arXiv:1112.0493 [hep-ph]} \BibitemShut {NoStop}%
\bibitem [{\citenamefont {Bernal}\ \emph {et~al.}(2016)\citenamefont {Bernal},
  \citenamefont {Chu}, \citenamefont {Garcia-Cely}, \citenamefont {Hambye},\
  and\ \citenamefont {Zaldivar}}]{Bernal:2015ova}%
  \BibitemOpen
  \bibfield  {author} {\bibinfo {author} {\bibfnamefont {N.}~\bibnamefont
  {Bernal}}, \bibinfo {author} {\bibfnamefont {X.}~\bibnamefont {Chu}},
  \bibinfo {author} {\bibfnamefont {C.}~\bibnamefont {Garcia-Cely}}, \bibinfo
  {author} {\bibfnamefont {T.}~\bibnamefont {Hambye}}, \ and\ \bibinfo {author}
  {\bibfnamefont {B.}~\bibnamefont {Zaldivar}},\ }\href {\doibase
  10.1088/1475-7516/2016/03/018} {\bibfield  {journal} {\bibinfo  {journal}
  {JCAP}\ }\textbf {\bibinfo {volume} {03}},\ \bibinfo {pages} {018} (\bibinfo
  {year} {2016})},\ \Eprint {http://arxiv.org/abs/1510.08063} {arXiv:1510.08063
  [hep-ph]} \BibitemShut {NoStop}%
\bibitem [{\citenamefont {Bernal}\ and\ \citenamefont
  {Chu}(2016)}]{Bernal:2015xba}%
  \BibitemOpen
  \bibfield  {author} {\bibinfo {author} {\bibfnamefont {N.}~\bibnamefont
  {Bernal}}\ and\ \bibinfo {author} {\bibfnamefont {X.}~\bibnamefont {Chu}},\
  }\href {\doibase 10.1088/1475-7516/2016/01/006} {\bibfield  {journal}
  {\bibinfo  {journal} {JCAP}\ }\textbf {\bibinfo {volume} {01}},\ \bibinfo
  {pages} {006} (\bibinfo {year} {2016})},\ \Eprint
  {http://arxiv.org/abs/1510.08527} {arXiv:1510.08527 [hep-ph]} \BibitemShut
  {NoStop}%
\bibitem [{\citenamefont {Feng}\ \emph {et~al.}(2008)\citenamefont {Feng},
  \citenamefont {Tu},\ and\ \citenamefont {Yu}}]{Feng:2008mu}%
  \BibitemOpen
  \bibfield  {author} {\bibinfo {author} {\bibfnamefont {J.~L.}\ \bibnamefont
  {Feng}}, \bibinfo {author} {\bibfnamefont {H.}~\bibnamefont {Tu}}, \ and\
  \bibinfo {author} {\bibfnamefont {H.-B.}\ \bibnamefont {Yu}},\ }\href
  {\doibase 10.1088/1475-7516/2008/10/043} {\bibfield  {journal} {\bibinfo
  {journal} {JCAP}\ }\textbf {\bibinfo {volume} {10}},\ \bibinfo {pages} {043}
  (\bibinfo {year} {2008})},\ \Eprint {http://arxiv.org/abs/0808.2318}
  {arXiv:0808.2318 [hep-ph]} \BibitemShut {NoStop}%
\bibitem [{\citenamefont {Ackerman}\ \emph {et~al.}(2008)\citenamefont
  {Ackerman}, \citenamefont {Buckley}, \citenamefont {Carroll},\ and\
  \citenamefont {Kamionkowski}}]{Ackerman:mha}%
  \BibitemOpen
  \bibfield  {author} {\bibinfo {author} {\bibfnamefont {L.}~\bibnamefont
  {Ackerman}}, \bibinfo {author} {\bibfnamefont {M.~R.}\ \bibnamefont
  {Buckley}}, \bibinfo {author} {\bibfnamefont {S.~M.}\ \bibnamefont
  {Carroll}}, \ and\ \bibinfo {author} {\bibfnamefont {M.}~\bibnamefont
  {Kamionkowski}},\ }\href {\doibase 10.1103/PhysRevD.79.023519} {\bibfield  {journal}
  {\bibinfo  {journal} {Phys. Rev. D}\ }\textbf {\bibinfo {volume} {79}},\ \bibinfo
  {pages} {023519} (\bibinfo {year} {2009})},\ \Eprint
  {http://arxiv.org/abs/0810.5126} {arXiv:0810.5126 [hep-ph]} \BibitemShut
  {NoStop}%
\bibitem [{\citenamefont {Hodges}(1993)}]{Hodges:1993yb}%
  \BibitemOpen
  \bibfield  {author} {\bibinfo {author} {\bibfnamefont {H.}~\bibnamefont
  {Hodges}},\ }\href {\doibase 10.1103/PhysRevD.47.456} {\bibfield  {journal}
  {\bibinfo  {journal} {Phys. Rev. D}\ }\textbf {\bibinfo {volume} {47}},\
  \bibinfo {pages} {456} (\bibinfo {year} {1993})}\BibitemShut {NoStop}%
\bibitem [{\citenamefont {Berezhiani}\ \emph {et~al.}(1996)\citenamefont
  {Berezhiani}, \citenamefont {Dolgov},\ and\ \citenamefont
  {Mohapatra}}]{Berezhiani:1995am}%
  \BibitemOpen
  \bibfield  {author} {\bibinfo {author} {\bibfnamefont {Z.}~\bibnamefont
  {Berezhiani}}, \bibinfo {author} {\bibfnamefont {A.}~\bibnamefont {Dolgov}},
  \ and\ \bibinfo {author} {\bibfnamefont {R.}~\bibnamefont {Mohapatra}},\
  }\href {\doibase 10.1016/0370-2693(96)00219-5} {\bibfield  {journal}
  {\bibinfo  {journal} {Phys. Lett. B}\ }\textbf {\bibinfo {volume} {375}},\
  \bibinfo {pages} {26} (\bibinfo {year} {1996})},\ \Eprint
  {http://arxiv.org/abs/hep-ph/9511221} {arXiv:hep-ph/9511221} \BibitemShut
  {NoStop}%
\bibitem [{\citenamefont {Bernal}\ \emph {et~al.}(2020)\citenamefont {Bernal},
  \citenamefont {Donini}, \citenamefont {Folgado},\ and\ \citenamefont
  {Rius}}]{Bernal:2020fvw}%
  \BibitemOpen
  \bibfield  {author} {\bibinfo {author} {\bibfnamefont {N.}~\bibnamefont
  {Bernal}}, \bibinfo {author} {\bibfnamefont {A.}~\bibnamefont {Donini}},
  \bibinfo {author} {\bibfnamefont {M.~G.}\ \bibnamefont {Folgado}}, \ and\
  \bibinfo {author} {\bibfnamefont {N.}~\bibnamefont {Rius}},\ }\href {\doibase
  10.1007/JHEP09(2020)142} {\bibfield  {journal} {\bibinfo  {journal} {JHEP}\
  }\textbf {\bibinfo {volume} {09}},\ \bibinfo {pages} {142} (\bibinfo {year}
  {2020})},\ \Eprint {http://arxiv.org/abs/2004.14403} {arXiv:2004.14403
  [hep-ph]} \BibitemShut {NoStop}%
\bibitem [{\citenamefont {Weldon}(1982)}]{Weldon:1982bn}%
  \BibitemOpen
  \bibfield  {author} {\bibinfo {author} {\bibfnamefont {H.}~\bibnamefont
  {Weldon}},\ }\href {\doibase 10.1103/PhysRevD.26.2789} {\bibfield  {journal}
  {\bibinfo  {journal} {Phys. Rev. D}\ }\textbf {\bibinfo {volume} {26}},\
  \bibinfo {pages} {2789} (\bibinfo {year} {1982})}\BibitemShut {NoStop}%
\bibitem [{\citenamefont {Quiros}(1999)}]{Quiros:1999jp}%
  \BibitemOpen
  \bibfield  {author} {\bibinfo {author} {\bibfnamefont {M.}~\bibnamefont
  {Quiros}},\ }in\ \href@noop {} {\emph {\bibinfo {booktitle} {{ICTP Summer
  School in High-Energy Physics and Cosmology}}}}\ (\bibinfo {year} {1999})\
  pp.\ \bibinfo {pages} {187--259},\ \Eprint
  {http://arxiv.org/abs/hep-ph/9901312} {arXiv:hep-ph/9901312} \BibitemShut
  {NoStop}%
\bibitem [{\citenamefont {Zyla}\ \emph {et~al.}(2020)\citenamefont {Zyla} \emph
  {et~al.}}]{Zyla:2020zbs}%
  \BibitemOpen
  \bibfield  {author} {\bibinfo {author} {\bibfnamefont {P.}~\bibnamefont
  {Zyla}} \emph {et~al.} (\bibinfo {collaboration} {Particle Data Group}),\
  }\href {\doibase 10.1093/ptep/ptaa104} {\bibfield  {journal} {\bibinfo
  {journal} {PTEP}\ }\textbf {\bibinfo {volume} {2020}},\ \bibinfo {pages}
  {083C01} (\bibinfo {year} {2020})}\BibitemShut {NoStop}%
\bibitem [{\citenamefont {Aghanim}\ \emph {et~al.}(2020)\citenamefont {Aghanim}
  \emph {et~al.}}]{Aghanim:2018eyx}%
  \BibitemOpen
  \bibfield  {author} {\bibinfo {author} {\bibfnamefont {N.}~\bibnamefont
  {Aghanim}} \emph {et~al.} (\bibinfo {collaboration} {Planck}),\ }\href
  {\doibase 10.1051/0004-6361/201833910} {\bibfield  {journal} {\bibinfo
  {journal} {Astron. Astrophys.}\ }\textbf {\bibinfo {volume} {641}},\ \bibinfo
  {pages} {A6} (\bibinfo {year} {2020})},\ \Eprint
  {http://arxiv.org/abs/1807.06209} {arXiv:1807.06209 [astro-ph.CO]}
  \BibitemShut {NoStop}%
\end{thebibliography}%

\end{document}